\documentclass[aps,a4paper,reprint,10pt,superscriptaddress,prl]{revtex4-2}
\usepackage[utf8]{inputenc}
\usepackage{amsmath}
\usepackage{amssymb}
\usepackage{braket}	
\usepackage{graphicx}
\usepackage{xcolor}
\usepackage{mathrsfs}
\usepackage[colorlinks=true, citecolor=.]{hyperref}
\usepackage{soul}
\usepackage[export]{adjustbox}
\usepackage{physics}
\usepackage{pgfplots}
\usepackage{tikz}

\newcommand{\R}{\mathbb{R}}
\newcommand{\qtr}{\operatorname{Tr}}
\renewcommand{\tr}{\operatorname{tr}}

\definecolor{mypink}{HTML}{C65B80}
\definecolor{myorange}{HTML}{C7733B}
\definecolor{mygreen}{HTML}{5F8340}
\definecolor{myblue}{HTML}{6C93C5}
\definecolor{mypurple}{HTML}{9765CA}

\begin{document}

\title{The Most Informative Cram\'er--Rao Bound for Quantum Two-Parameter Estimation with Pure State Probes}
\author{Simon K. Yung} 
\email{sksyung@gmail.com}
\affiliation{Centre for Quantum Computation and Communication Technology, Department of Quantum Science and Technology, Research School of Physics, The Australian National University, Canberra, ACT 2601, Australia.}
\author{C. M. Yung}
\noaffiliation
\author{Lorc\'an O. Conlon}
\affiliation{Joint Quantum Institute and Joint Center for Quantum Information and Computer Science, NIST/University of Maryland, College Park, Maryland 20742, USA.}
\author{Syed M. Assad}
\affiliation{A*STAR Quantum Innovation Centre (Q.INC), Agency for Science, Technology and Research (A*STAR), 2 Fusionopolis Way, Innovis, 138634, Singapore.}
\date{\today}

\begin{abstract}
	Optimal measurements for quantum multiparameter estimation are complicated by the uncertainty principle. Generally, there is a trade-off between the precision with which different parameters can be simultaneously estimated. The task of determining the minimum achievable estimation error is a central task of multiparameter quantum metrology. For estimating parameters encoded in pure quantum states, the ultimate limit is known, but is given by the solution of a non-trivial minimisation problem. We present a new expression for the achievable bound for two-parameter estimation with pure states that is considerably simpler. We also determine the optimal measurements, completing the problem of two-parameter estimation with pure state probes. To demonstrate the utility of our result, we determine the precision limit for estimating displacements using grid states.  
\end{abstract}

\maketitle

\textit{Introduction.}---Progress in quantum technologies is driven by an improved understanding of the potentials and limitations of quantum mechanics, and how these boundaries can be reached. Advances in quantum metrology have enabled enhanced sensitivity in applications such as gravitational wave observatories~\cite{aasi_enhanced_2013} and biological imaging~\cite{casacio_quantum-enhanced_2021}. From a fundamental perspective, quantum metrology is grounded in the theory of quantum parameter estimation, which is concerned with the precise estimation of parameters encoded in quantum states.  

Underpinned by the pioneering work of Helstrom~\cite{helstrom_minimum_1967,helstrom_minimum_1968,helstrom_quantum_1976} and Holevo~\cite{holevo_statistical_1973,holevo_noncommutative_1976,holevo_probabilistic_2011}, multiparameter quantum estimation theory has experienced rapid growth and progression~\cite{liu_quantum_2019,albarelli_perspective_2020,demkowicz-dobrzanski_multi-parameter_2020,sidhu_geometric_2020}. This is evident in the variety of theoretical tools that can now be used to analyse the accessible information in multiparameter systems~\cite{albarelli_evaluating_2019,conlon_efficient_2021,lu_incorporating_2021,hayashi_tight_2023,yung_comparison_2024,yung_saturating_2025}, and the numerous works in which fundamental theory has guided demonstrations of quantum enhancement in multiparameter systems~\cite{hou_deterministic_2018,conlon_approaching_2023,conlon_discriminating_2023,zhou_experimental_2025}.  

Despite this progress, generally applicable and easy-to-calculate expressions for the minimum achievable estimation error have remained elusive, with the exception of special cases, such as in two-dimensional Hilbert spaces~\cite{gill_state_2000,nagaoka_new_2005,suzuki_explicit_2016}. Fundamentally, this stems from the complexity of the problem, as one must a priori optimise over all possible measurements. Increased access to understanding of the attainable precision and the limitations of incompatibility will be beneficial towards enabling further progress in the field and its applications.   

A common approach to determining the attainable estimation error is to derive a lower bound (that is not necessarily attainable, but is tractable to compute), and then prove that the bound is achievable under certain conditions. Matsumoto~\cite{matsumoto_new_2002} called this the ``indirect approach'', which has been applied for the Holevo Cram\'er--Rao bound~\cite{holevo_statistical_1973,holevo_noncommutative_1976,holevo_probabilistic_2011} and Nagaoka Cram\'er--Rao bound~\cite{nagaoka_new_2005,nagaoka_generalization_2005}. A drawback with this approach is that, even when attainability is proven, the optimal measurements are not immediately determined and further optimisation is required to find them.

On the other hand, in a ``direct approach'', one can convert the minimisation over all measurements into an alternative tractable minimisation problem, possibly by imposing certain conditions before the conversion. Direct approaches have been taken by Gill and Massar~\cite{gill_state_2000} and Matsumoto~\cite{matsumoto_new_2002}. 

In this work, we take a direct approach and derive an expression for the attainable estimation error for quantum two-parameter estimation with pure state probes, in any Hilbert space dimension. Our expression is significantly simpler than alternative general results. A by-product of our approach is that the optimal measurement is fully determined. In this regard, two-parameter quantum estimation with pure states can be considered solved. For mixed states, our result provides a simple lower bound to the mean squared error, though it is not in general attainable.

\textit{Quantum parameter estimation and Cram\'er--Rao bounds.}---For two-parameter estimation with pure state probes, we consider a smooth family of quantum states $\{\ket{\psi_\theta}\in \mathcal{H}_d\ | \ \theta=(\theta_1,\theta_2)\in \R^2 \}$, where $d=\dim\mathcal{H}_d$ is the dimension of the Hilbert space. The values of the parameters can be estimated based on the results of measurements performed on a large number $N$ of identical copies of the quantum state. A general measurement can be described by a positive operator-valued measure (POVM) $\Pi = \{\Pi_i  | \sum_i \Pi_i =\openone_d, \Pi_i \succeq 0\}$, where the conditional probability of the outcomes are given by the Born rule $p(i|\theta) = \bra{\psi_\theta}\Pi_i\ket{\psi_\theta}$. The statistical results can be post-processed with an estimator, $\hat{\theta}$, to obtain an estimate of the parameters. 

It is customary to quantify the estimation performance by the estimator's mean squared error matrix, $V(\hat\theta,\Pi)$, with elements $V(\hat\theta,\Pi)_{jk} = \mathbb{E}[(\hat\theta_j-\theta_j)(\hat\theta_k-\theta_k)]$, where the expectation value is taken over the probability distribution $\{p(i|\theta)\}$. We assume that the true value of the parameters is approximately known, and that the estimator is locally unbiased around this value so that $\mathbb{E}[\hat\theta]=\theta$ and $\partial\mathbb{E}[\hat\theta_j]/\partial\theta_k = \delta_{jk}$. Under these conditions, the mean squared error is bounded below as $V(\hat\theta,\Pi)\succeq F(\Pi)^{-1}/N$, where $F(\Pi)$ is the classical Fisher information of $\{p(i|\theta)\}$~\cite{fisher_mathematical_1922,cramer_mathematical_1946,rao_minimum_1947}. This lower bound can be attained in the limit of a large $N$, e.g., by maximum likelihood estimation, and we therefore suppress the dependence of $\hat{\theta}$ in the mean squared error. We also absorb the factor of $N$ into the mean squared error. 

The central task of theoretical quantum parameter estimation is to determine the mean squared errors that can be attained by measurements. That is, how precisely can we determine the value of $\theta$? Several lower bounds on the mean squared error have been developed, beginning with the work of Helstrom~\cite{helstrom_minimum_1967,helstrom_minimum_1968,helstrom_quantum_1976}. Below, we outline the major developments, with a particular emphasis on pure state estimation. 

In analogy to the classical Fisher information, Helstrom developed the quantum Fisher information~\cite{helstrom_minimum_1967,helstrom_minimum_1968,helstrom_quantum_1976}, given by $J_{jk} = \Re\qtr[\rho L_jL_k]=\qtr[\rho\{L_j,L_k\}]/2$, where $\rho=\ketbra{\psi_\theta}$ is the density operator corresponding to $\ket{\psi_\theta}$ and $L_j$ is the symmetric logarithmic derivative (SLD), a Hermitian operator satisfying $\partial\rho/\partial\theta_j = (\rho L_j+L_j\rho)/2$. For convenience later, here we define the corresponding imaginary part $\tilde{J}=\Im \qtr[\rho L_jL_k]=\qtr[\rho[L_j,L_k]]/(2i) $. For pure states (or indeed any rank-deficient density operator), the SLD operators are not unique. However, Fujiwara and Nagaoka demonstrated that the quantum Fisher information (and $\tilde{J}$) are nevertheless uniquely defined~\cite{fujiwara_quantum_1995}. The quantum Fisher information sets a limit on the mean squared error $V\succeq J^{-1}$, called the quantum Cram\'er--Rao bound~\cite{braunstein_statistical_1994,paris_quantum_2009}, a consequence of the fact that $J$ bounds the classical Fisher information of any measurement. Due to non-commutativity of the optimal measurements for each parameter, this lower bound is generally not attainable for multiparameter estimation, and alternatives must be sought to gain more insight into the attainable mean squared error~\cite{liu_quantum_2019,albarelli_perspective_2020,demkowicz-dobrzanski_multi-parameter_2020,sidhu_geometric_2020,ragy_compatibility_2016}. 

It is typical to attempt to minimise the scalar quantity $\tr[WV(\Pi)]$, for a symmetric positive-definite matrix $W$, called the weight matrix \footnote{We use ``$\tr$'' to denote a trace over the classical parameter space, to distinguish the trace over the quantum Hilbert space ($\qtr$)}. The weight matrix can represent the quadratic component of an arbitrary cost function $h(\hat\theta,\theta)$, which is the dominant term in the asymptotic error~\cite{gill_state_2000}. The field has largely turned to developing lower bounds for $\tr[WV(\Pi)]$, which are usually called ``Cram\'er--Rao-type bounds''. With this in mind, the quantity that is ultimately of interest is
\begin{equation}
	\mathcal{C}^\text{MI}(W) = \min_\Pi\{\tr[WV(\Pi)]\} = \min_\Pi\{\tr[WF(\Pi)^{-1}]\}.
\end{equation} 
Following Nagaoka~\cite{nagaoka_new_2005}, we call this the ``most informative Cram\'er--Rao bound'', which represents the minimum attainable mean squared error. Nagaoka gave a closed-form expression for $\mathcal{C}^\text{MI}$ when $d=2$~\cite{nagaoka_new_2005,nagaoka_generalization_2005}.

Taking a different approach, Gill and Massar~\cite{gill_state_2000} showed that given a measurement $\Pi$, the associated classical Fisher information $F(\Pi)$ satisfies the inequality:
\begin{equation}
	\tr[J^{-1}F(\Pi)]\geq (d-1). \label{eq:GMineq}
\end{equation}
By minimising $\tr[WF^{-1}]$ subject to 
\begin{equation}
	\tr[J^{-1}F]\geq (d-1) \label{eq:GMtarget}
\end{equation}
using Lagrange multipliers, they derived a Cram\'er--Rao-type bound, which coincides with Nagaoka's result for $d=2$. Gill and Massar called the positive semidefinite matrix $F$ in 
Eq.~\eqref{eq:GMtarget} a ``target Fisher information'' (ie. not necessarily backed by a measurement). They showed that in $d=2$, every target Fisher information matrix is indeed the classical Fisher information of a measurement on the system, thus demonstrating attainability of their bound. However, the same is not true when $d>2$.  

Another important lower bound was developed by Holevo and is commonly known as the Holevo Cram\'er--Rao bound (HCRB)~\cite{holevo_noncommutative_1976,holevo_probabilistic_2011}. It is the solution of a minimisation problem over $d\times d$ Hermitian matrices. In general, the HCRB is attainable with a collective measurement performed jointly on an asymptotically large number of copies of the state~\cite{kahn_local_2009,yamagata_quantum_2013,yang_attaining_2019}. However, for pure states it is attainable with separable measurements~\cite{matsumoto_new_2002}. That is, for pure states, the HCRB is equal to $\mathcal{C}^\text{MI}$ and there is no advantage to be gained with entanglement at the measurement stage. While it was shown that the HCRB can be numerically evaluated using a semidefinite program~\cite{albarelli_evaluating_2019}, this does not determine the optimal measurement, which thus requires further optimisation. 

\textit{Tight inequality for the Fisher information matrix.}---We present an inequality for the classical Fisher information of any measurement, for which equality is possible with certain measurements. The functional reason for seeking an inequality for the Fisher information is that it enables a direct approach to calculating the attainable mean squared error.

The inequality is equivalent to a relation presented in Ref.~\cite{matsumoto_new_2002}. It can also be derived by modifying Lu and Wang's information regret trade-off~\cite{lu_incorporating_2021}, as was done recently to obtain a similar relation~\cite{hu_quantum_2025}. Nevertheless we outline here a derivation that uses Branciard's geometric inequality~\cite{branciard_error-tradeoff_2013}, as it is instructive to understand the source of the inequality, and it will be helpful in discussing optimal measurements. Additional details are provided in the Supplemental Material.

The basis for the inequality is the approximate joint measurement of the (generally) non-commuting observables, $L_1$, $L_2$, that are individually optimal for estimating $\theta_1$ and $\theta_2$. Any POVM can be expressed as a projective measurement on an extended Hilbert space, i.e., a measurement $\{\ketbra{m}\}$ made on the state $\ket{\psi_\theta}\ket{\xi}$ where $\ket{\xi}\in\mathcal{H}_a$ is a state of an ancillary system. The approximate joint measurement is then described as the simultaneous measurement of commuting observables $O_1$, $O_2$, in an eigenbasis $\{\ket{m}\}$. Following Wang \textit{et al.}~\cite{wang_tight_2025}, we consider the states
\begin{align}
	\ket{l_j} &= (L_j\otimes \openone_a)\ket{\psi_\theta}\ket{\xi}, \quad 
	\ket{o_j} = O_j\ket{\psi_\theta}\ket{\xi}. 
\end{align}
The inner products between these vectors are related to the quantum ($\ket{l_j}$) and classical ($\ket{o_j}$) Fisher information matrices. We apply Branciard's geometric inequality~\cite{branciard_error-tradeoff_2013} to the differences $\ket{o_j}-\ket{l_j}$, which quantifies that (in general) the differences cannot simultaneously vanish. This gives rise to an inequality involving only the diagonal elements of the Fisher information matrices, equivalent to Lu and Wang's information regret trade-off relation (IRTR)~\cite{lu_incorporating_2021}. 

To account for the off-diagonal elements of the Fisher information, which are responsible for quantifying correlations between parameters, we resort to reparametrisation of the system, as done by Hu and Lu~\cite{hu_quantum_2025} to extend the validity~\cite{yung_comparison_2024} of the IRTR. Namely, by reparameterising the quantum state via a transformation described by a Jacobian $\partial\theta_i/\partial\theta'_j = J^{-1/2}Q$, with $Q$ an orthogonal matrix, the Fisher information matrices become
\begin{align}
\begin{split}
	J' &= (J^{-1/2}Q)^\top J(J^{-1/2}Q) = I, \\ 
	F' &= (J^{-1/2}Q)^\top F(J^{-1/2}Q).
	\end{split}\label{eq:transfFIM}
\end{align}
With an appropriate choice of matrix $Q$, both the transformed quantum and classical Fisher information matrices become diagonal. Hu and Lu referred to this particular parameterisation as the ``canonical parameterisation'' of the problem. 

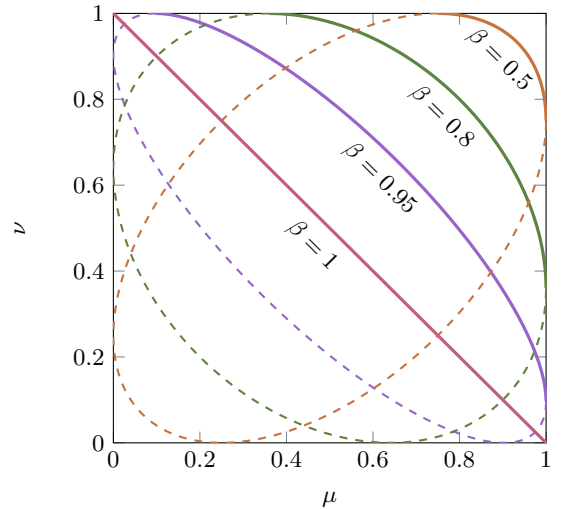
\begin{figure}
	\centering
	\begin{tikzpicture}
  	\begin{axis}[
    xlabel={$\mu$},
    ylabel={$\nu$},
    xmin=0, xmax=1,
    ymin=0, ymax=1,
    axis equal image,
  ]
  		\pgfmathsetmacro{\bet}{0.95}
  		\pgfmathsetmacro{\eta}{asin(\bet)/2}
  
  		\addplot [domain=-\eta:\eta, samples=50,very thick,mypurple] 
    		({(cos(x + \eta))^2}, {(cos(x-\eta))^2});
  		\addplot [domain=\eta:180-\eta, samples=150, dashed,thick,mypurple] 
    		({(cos(x + \eta))^2}, {(cos(x-\eta))^2});
    	\pgfmathsetmacro{\centercoordone}{(cos(\eta))^2}
  		\node[anchor=north, inner sep=5pt,rotate=-45] at (axis cs:\centercoordone,\centercoordone) {\small $\beta = 0.95$};
   
  		\pgfmathsetmacro{\bet}{0.8}
  		\pgfmathsetmacro{\eta}{asin(\bet)/2}
  		\addplot [domain=-\eta:\eta, samples=50,very thick,mygreen] 
    		({(cos(x + \eta))^2}, {(cos(x-\eta))^2});
  		\addplot [domain=\eta:180-\eta, samples=150, dashed,thick,mygreen] 
    		({(cos(x + \eta))^2}, {(cos(x-\eta))^2});
    	\pgfmathsetmacro{\centercoordtwo}{(cos(\eta))^2}
    	\node[anchor=north, inner sep=5pt,rotate=-45] at (axis cs:\centercoordtwo,\centercoordtwo) {\small $\beta = 0.8$};
  
  		\pgfmathsetmacro{\bet}{0.5}
  		\pgfmathsetmacro{\eta}{asin(\bet)/2}  
  		\addplot [domain=-\eta:\eta, samples=50,very thick,myorange] 
    		({(cos(x + \eta))^2}, {(cos(x-\eta))^2});
  		\addplot [domain=\eta:180-\eta, samples=150, dashed,thick,myorange] 
    		({(cos(x + \eta))^2}, {(cos(x-\eta))^2});
		\pgfmathsetmacro{\centercoordthr}{(cos(\eta))^2}
    	\node[anchor=north, inner sep=5pt,rotate=-45] at (axis cs:\centercoordthr,\centercoordthr) {\small $\beta = 0.5$};
    	
    	\pgfmathsetmacro{\bet}{1}
  		\pgfmathsetmacro{\eta}{asin(\bet)/2}  
  		\addplot [domain=-\eta:\eta, samples=2, very thick,mypink] 
    		({(cos(x + \eta))^2}, {(cos(x-\eta))^2});
		\pgfmathsetmacro{\centercoordfour}{(cos(\eta))^2}
    	\node[anchor=north, inner sep=5pt,rotate=-45] at (axis cs:\centercoordfour,\centercoordfour) {\small $\beta = 1$};
    	
  		\end{axis}

	\end{tikzpicture}

	\caption{Allowed regions for the eigenvalues $\mu$, $\nu$ of the matrix $G(\Pi)$, for sample values of $\beta$. The solid lines represent the boundaries of the allowed regions (above the solid lines is not possible, as are values less than 0 or greater than 1). The dashed curves demonstrate that the boundary of the regions is an arc of an ellipse. For $\beta=1$, the ellipse degenerates to a line.}
	\label{fig:ellipsereg}
\end{figure}

Applying the aforementioned inequality in the canonical parametrisation, and then reverting back to the original parametrisation, we obtain an inequality for the classical Fisher information of any measurement
\begin{equation}
	\sqrt{\det[G(\Pi)]}-\sqrt{\det[I-G(\Pi)]}\leq \sqrt{1-\beta^2}, \label{eq:newinequality}
\end{equation}
where $G(\Pi)= J^{-1/2}F(\Pi)J^{-1/2}$. Here, $\beta$ is the absolute value of the eigenvalues of the matrix $J^{-1}\tilde{J}$ (which has eigenvalues $\pm i\beta$ with $0\leq \beta\leq 1$~\cite{matsumoto_new_2002}). The inequality can be simplified and expressed in terms of the eigenvalues, $\mu$ and $\nu$, of $G(\Pi)$, as
\begin{equation}
	\sqrt{\mu\nu}-\sqrt{(1-\mu)(1-\nu)}\leq \sqrt{1-\beta^2}. \label{eq:newinequalityeig}
\end{equation}
The eigenvalues must also satisfy $0\leq \mu,\nu\leq 1$, from the inequality $0\preceq F(\Pi)\preceq J$. This constrains the eigenvalues to a region bounded by the unit square and the arc of an ellipse, as depicted in Fig.~\ref{fig:ellipsereg}. The arc of the ellipse is exactly the region corresponding to saturating the inequality of Eq.~\eqref{eq:newinequality}, and can be parameterised by
\begin{equation}
	\mu = \cos^2(\varphi-\eta), \ \nu=\cos^2(\varphi+\eta), \ \eta = \arcsin(\beta)/2, \label{eq:parameterisedeig}
\end{equation}
with $-\eta\leq \varphi\leq \eta$. Intuitively, these eigenvalues represent the information about the canonical parameters obtained by a measurement. 

Crucially, equality in Eq.~\eqref{eq:newinequality} is not only attainable, but a measurement exists corresponding to any $G$ saturating the inequality. When $\beta=1$, the vectors $\ket{l_i}$ are linearly dependent and the problem is contained within the two-dimensional subspace $\operatorname{Span}\{\ket{\psi_\theta},\ket{l_1},\ket{l_2}\}$. Within this subspace, we construct a four-outcome POVM that depends on $\varphi$ and saturates the inequality with the eigenvalues in Eq.~\eqref{eq:parameterisedeig} (see Supplemental Materials for details). Such a measurement can then be adapted to obtain any $G$ with those eigenvalues, equivalent to selecting a different $Q$ in Eq.~\eqref{eq:transfFIM}. Alternatively, the measurement proposed in Ref.~\cite{wang_tight_2025} is also optimal and has three outcomes. For $\beta<1$, the relevant subspace is three-dimensional, and we can similarly construct an optimal POVM with three outcomes (see Supplemental Materials for details). The measurement proposed in Ref.~\cite{branciard_error-tradeoff_2013} is also optimal.

\textit{The most informative Cram\'er--Rao bound.}---We now proceed to derive an expression for the most informative bound for two-parameter estimation with pure states. Our approach is most similar to Gill and Massar's, with the main difference that we use the inequality in Eq.~\eqref{eq:newinequality} (which is equivalent to Eq.~\eqref{eq:GMineq} when $d=2$). 

The most informative Cram\'er--Rao bound, $\mathcal{C}^\text{MI}(W)$, is the minimum of $\tr[WF(\Pi)^{-1}]$ over all measurements $\Pi$
\begin{equation}
	\mathcal{C}^\text{MI}(W) = \min_{\Pi \text{ POVM}}\tr[WF(\Pi)^{-1}] 
\end{equation}
Given a weight matrix $W$ we define $S=J^{-1/2}W J^{-1/2}$, which is also positive semidefinite, and let its ordered eigenvalues be denoted by $\lambda_1(S)\geq \lambda_2(S)$. For a measurement $\Pi$ we then have
\begin{equation}
    \tr[W F(\Pi)^{-1}] = \tr[S G(\Pi)^{-1}] \geq \frac{\lambda_1(S)}{\mu} + \frac{\lambda_2(S)}{\nu},\label{eq:vonneumann}
\end{equation}
where $\mu\geq \nu$ are the eigenvalues of $G(\Pi)$. The inequality above is a special case of Ruhe's inequality for the trace of a product of positive semidefinite matrices~\cite{marshall_inequalities_2011}. Equality holds if and only if $S$ and $G^{-1}$ commute. Notice that the numerators are the summands of $\tr[S]=\tr[WJ^{-1}]$, the scalar quantum Cram\'er--Rao bound. The right-hand side of \eqref{eq:vonneumann} is a decreasing function of $\mu$ and $\nu$, so the minimum occurs somewhere on the arc of the ellipse defined by equality in Eq.~\eqref{eq:newinequalityeig}, with eigenvalues of $G(\Pi)$ parameterised as in Eq.~\eqref{eq:parameterisedeig}. As described earlier and shown in the Supplemental Material, there exists a measurement $\Pi$ giving rise to any $G(\Pi)$ with eigenvalues $\mu,\nu$ on the arc of the ellipse. In particular, a $G(\Pi)$ which commutes with $S$, ensuring Ruhe's inequality above is saturated. 

Therefore, we have the general result for the most informative Cram\'er--Rao bound:
\begin{equation}
	\mathcal{C}^\text{MI}(W) = \min_{0\leq\varphi\leq \eta} \left\{\frac{\lambda_1(S)}{\cos^2(\varphi-\eta)}+ \frac{\lambda_2(S)}{\cos^2(\varphi+\eta)} \right\},  \label{eq:CMInew} 
\end{equation}
with $\eta = \arcsin(\beta)/2$. By changing the minimisation domain to $\varphi\in[-\eta,\eta]$, the numerators in Eq.~\eqref{eq:CMInew} can be replaced by the eigenvalues of $S$ in any order. The solution to the minimisation, $\varphi^\star$, determines the optimal measurement, as described in the Supplemental Material. We note that the minimisation over $\varphi$ can also be reduced to finding a zero of a fourth-order polynomial in $x=\tan\varphi$. A closed-form expression for $\mathcal{C}^\text{MI}(W)$ therefore exists, although it is unwieldy (see Supplemental Material for details on the conversion). 

\textit{Connection to previous results and special cases.}---For certain special cases, from Eq.~\eqref{eq:CMInew} we can recover known results for the attainable bound. For instance, for $\beta=0$, we have $\eta=0$ and the eigenvalues of $G$ can be simultaneously maximised, so $\mathcal{C}^\text{MI}(W)= \tr[S]= \mathcal{C}^\text{SLD}(W)$. This corresponds to the case where there is no parameter incompatibility and the SLD operators weakly commute, $\bra{\psi}[L_1,L_2]\ket{\psi}=0$, where it is known that the SLD Cram\'er--Rao bound is attainable~\cite{ragy_compatibility_2016}. 

On the other hand, when the incompatibility is maximal, $\beta=1$, we recover Gill and Massar's formula~\cite{gill_state_2000}
\begin{equation}
	\mathcal{C}^\text{MI}(W;\beta=1) = \left(\tr\left[\sqrt{J^{-1/2}WJ^{-1/2}}\right]\right)^2,
\end{equation}
which now holds in any Hilbert space dimension. This coincides with the exact results for qubits~\cite{gill_state_2000,nagaoka_new_2005,suzuki_explicit_2016}, because for $d=2$, we have $\beta=1$ (See Supplemental Material for a simple demonstration).

For $\lambda_1(S)=\lambda_2(S)=s$, i.e., when $W$ is proportional to $J$, the minimum occurs at $\varphi=0$ and we recover Matsumoto's result~\cite{matsumoto_new_2002}
\begin{equation}
	\mathcal{C}^\text{MI}(W\propto J) = \frac{4s}{1+\sqrt{1-\beta^2}}.
\end{equation}

Lastly, we remark that an expression similar to Eq.~\eqref{eq:CMInew} for the HCRB was derived for a specific system in Ref.~\cite{gardner_achieving_2024}, where the HCRB was determined following the method of Ref.~\cite{bradshaw_tight_2017}, involving directly satisfying the constraints of the HCRB minimisation problem. 

\textit{Displacement sensing with grid states.}---To demonstrate the simplicity of calculating the achievable lower bound, we consider simultaneously estimating orthogonal displacements in phase space, encoded by the unitary $D(u,v) = e^{-iu\hat{p}+iv\hat{q}}$, where $\hat{q}$ and $\hat{p}$ are the position and momentum quadrature operators, satisfying $[\hat{q},\hat{p}]=i\hbar$. Grid states, superpositions of displaced squeezed states forming a grid in phase space (originally designed for quantum error correction~\cite{gottesman_encoding_2001}), enable simultaneous displacement sensing beyond the standard quantum limit~\cite{duivenvoorden_single-mode_2017}. The advantage of these non-Gaussian states has recently been demonstrated in a trapped-ion system~\cite{valahu_quantum-enhanced_2025}. 

\begin{figure}
    \centering
    \includegraphics{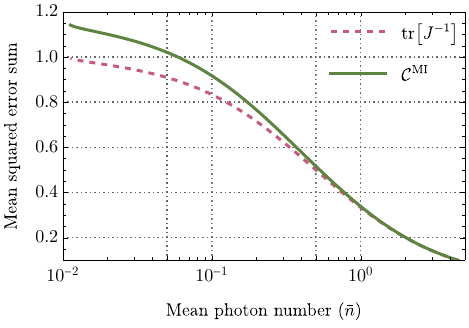}
    \caption{Mean squared error lower bounds for displacement sensing with grid states. The lower bounds and mean photon number are calculated numerically for different squeezing levels. }
    \label{fig:gridstate}
\end{figure}

Ref.~\cite{duivenvoorden_single-mode_2017} studied the quantum Cram\'er--Rao bound for estimating displacements with grid states \footnote{Strictly speaking, Ref.~\cite{duivenvoorden_single-mode_2017} derived a lower bound to the quantum Cram\'er--Rao bound. They determined $\tr[J]$ and argued that $\tr[J^{-1}]$ is minimised when $J$ is proportional to the identity. Numerically, we find that $J$ is close to proportional to the identity, but not quite (owing to the asymmetry between the $q$- and $p$-wavefunctions), so there is a small difference between $\tr[J^{-1}]$ and the result in Ref.~\cite{duivenvoorden_single-mode_2017}.}, but it remains to determine how close this is to the ultimate attainable bound. We find that the quantum Cram\'er--Rao bound is attainable in the limit that the mean photon number ($\bar{n}$) of the grid state goes to infinity, corresponding to infinite squeezing: $\beta \rightarrow 0$ as $\bar{n}\rightarrow \infty$. For small values of $\bar{n}$, there is a small gap, as shown in Fig.~\ref{fig:gridstate}. Notably, with our method there is only a small computation required to extend from the quantum Cram\'er--Rao bound to the ultimate atttainable bound. This allows for an accurate assessment of the optimality of experimentally accessible measurements. Details on the computation are provided in the Supplemental Material.

\textit{Discussion and conclusions.}---The major contribution of our work is a simple expression for the most informative Cram\'er--Rao bound for two-parameter estimation with pure states. It is substantially simpler than the alternative, the HCRB, indicating a practical advantage. In fact, the quantum Cram\'er--Rao bound is sometimes used as a proxy for the HCRB due to its simplicity and the fact that it is loose by at most a factor of $1+\beta\leq 2$~\cite{carollo_quantumness_2019}. This can, for example, simplify the search for an optimal probe state for sensing, but such an approach discards any information about the measurement incompatibility. In this regard, our result significantly reduces the barrier to accessing the full information about the attainable precisions. 

In addition, in contrast to the HCRB, the computational complexity of our minimisation result is independent of the dimension of the Hilbert space. In particular, this makes our result well-suited to studies of infinite-dimensional continuous-variable systems. 
In this way, our work enables detailed analyses of the fundamental precision limits in applications such as sensing with non-Gaussian states, as demonstrated above.  

As mentioned earlier, a by-product of our approach is that the optimal measurement is fully determined by the same minimisation as required to compute the attainable precision. This result is significant because determining the optimal measurement, in general, requires optimising over all POVMs. Here, the minimisation is not only significantly simplified, but the number of POVM elements is also known. For $\beta<1$, a projective measurement suffices, while for $\beta=1$, a non-projective measurement is required, utilising an ancillary system. 

Besides the computational advantage, our expression for $\mathcal{C}^\text{MI}$ offers several insights into the limit of quantum estimation with pure states, which are otherwise obscured in the constraints of the HCRB minimisation problem. For example, the dimension of the Hilbert space does not play a role, aside from precluding certain values of $\beta$, as is the case for $d=2$. Our result also demonstrates that estimation problems with the same $J$ and $\tilde{J}$ are equivalent. In fact, there is a stronger equivalence: any two problems with the same $\beta$ are isomorphic. 

Our expression also makes clear the role of $\beta$ in quantifying the incompatibility between the parameters. As $\beta$ increases, the eigenvalues of $G$ become more restricted, as seen in Fig.~\ref{fig:ellipsereg}. This is further reflected in the minimisation in Eq.~\eqref{eq:CMInew}, where a non-zero $\beta$ (and thus $\eta$) prohibits the simultaneous maximisation of the denominators. 

For mixed states, Eq.~\eqref{eq:CMInew} determines a lower bound for the mean squared error trace, but it is not generally attainable. This is true because the inequality also holds for mixed states, since any mixed state can be purified by treating it as the partial trace of a pure state in an extended Hilbert space~\cite{nielsen_quantum_2010,fujiwara_fibre_2008,kolodynski_efficient_2013,kolodynski_precision_2014}, and the purification can be chosen such that $J$ and $\beta$ are the same for the mixed and purified states (see Supplemental Materials for details). However, the attainability does not follow similarly, because, as discussed in Ref.~\cite{branciard_deriving_2014}, the measurement theoretically required to saturate the inequality may rely on the extended Hilbert space, but an experimenter only has access to the original space. 

We note that Ozawa~\cite{ozawa_error-disturbance_2014} strengthened Branciard's error trade-off relation for mixed states by replacing the incompatibility coefficient of the form $\tr[\rho[A,B]]$ (similar to $\beta$) with $\tr|\sqrt{\rho}[A,B]\sqrt{\rho}|$, where for an operator $|C|=\sqrt{C^\dagger C}$. This had the effect of tightening the error trade-off relation, though it is still not generally attainable. As such, while similar considerations could help tighten our lower bound for mixed states, it is not immediately helpful for calculating the most informative Cram\'er--Rao bound.  

Finally, we remark that for systems with several parameters, the canonical parameterisation is such that $\tilde{J}$ is block diagonal of the form
\begin{equation}
	\tilde{J} = \left(\begin{array}{@{}c|c@{}|c@{}}
		\begin{matrix}
			& \beta_1 \\
			-\beta_1 & 	
		\end{matrix} & & \\
		\hline
		& \begin{matrix}
			&\beta_2\  \\
			-\beta_2 
		\end{matrix} & \\
		\hline 
		& & \ \ddots
	\end{array}
	\right),
\end{equation}
where blank entries are zero, and $0\leq \beta_i\leq 1$~\cite{matsumoto_geometrical_1997}. In this way, in the canonical parameterisation, the system is decoupled into a series of two-parameter sub-models. In light of the direct role of $\beta$ in incompatibility, this suggests that, at least for pure states, quantum parameter incompatibility exists only between pairs of parameters, and not simultaneously between several parameters. We leave a detailed analysis of the most informative Cram\'er--Rao bound for several parameters to future work. 

We have taken a direct approach to determining the fundamental limit of the estimation error of two-parameter estimation with pure states. In doing so, we arrive at a simple expression for the attainable precision and corresponding optimal measurement. Our result provides insights into incompatibility in quantum parameter estimation, and enables simplified access to information about such incompatibility for practical applications.
 
\textit{Acknowledgements}---This research was supported by the Australian Research Council Centre of Excellence CE170100012. S.K.Y. is supported by the Australian Government Research Training Program Scholarship.

\bibliography{bib.bib}
\end{document}


\title{Supplemental Material for\\ \textit{The Most Informative Cram\'er--Rao Bound for Quantum Two-Parameter Estimation with Pure State Probes}}
\author{Simon K. Yung} 
\email{sksyung@gmail.com}
\affiliation{Centre for Quantum Computation and Communication Technology, Department of Quantum Science and Technology, Research School of Physics, The Australian National University, Canberra, ACT 2601, Australia.}
\author{C. M. Yung}
\noaffiliation
\author{Lorc\'an O. Conlon}
\affiliation{Joint Quantum Institute and Joint Center for Quantum Information and Computer Science, NIST/University of Maryland, College Park, Maryland 20742, USA.}
\author{Syed M. Assad}
\affiliation{A*STAR Quantum Innovation Centre (Q.INC), Agency for Science, Technology and Research (A*STAR), 2 Fusionopolis Way, Innovis, 138634, Singapore.}
\date{\today}

\maketitle

\setcounter{equation}{0}
\setcounter{figure}{0}
\setcounter{table}{0}
\setcounter{page}{1}
\makeatletter
\renewcommand{\theequation}{S\arabic{equation}}
\renewcommand{\thefigure}{S\arabic{figure}}
\renewcommand{\bibnumfmt}[1]{[S#1]}
\renewcommand{\citenumfont}[1]{S#1}

\tableofcontents

\section{Detailed derivation of the Fisher information inequality}
Here, we derive an inequality for the classical Fisher information of any measurement for jointly estimating $\theta_1$ and $\theta_2$. The commuting observables $O_1$, $O_2$, which act on the state $\ket{\psi_\theta}\ket{\xi}$, form an approximate joint measurement of the SLD operators $L_1$, $L_2$. As in Wang \textit{et al.}~\cite{wang_tight_2025}, we consider the states
\begin{equation}
	\ket{l_j} = (L_j\otimes \openone)\ket{\psi_\theta}\ket{\xi},\quad \ket{o_j} = O_j\ket{\psi_\theta}\ket{\xi}. 
\end{equation} 
The measurement operators can be expressed in the measurement basis $\{\ket{m}\}$ (e.g. defined by a POVM) as $O_j = \sum_m f_j(m)\ketbra{m}$. The optimal values of $f_j(m)$ (in the sense of minimising the squared approximation error, $\bra{\xi}\bra{\psi_\theta}(L_j\otimes\openone-O_j)^2\ket{\psi_\theta}\ket{\xi} $) are $f_j(m)= \partial_jp(m)/p(m)$, where $p(m)= \left|\braket{m}{\psi_\theta}\ket{\xi}\right|^2$ are the outcome probabilities of the measurement (if $p(m')=0$, set $f_j(m')=0$) \cite{wang_tight_2025}. This choice amounts to the choice of a classical estimator function given a POVM.  

We can relate the inner products between these states to matrix elements of $J$, $\tilde{J}$ and $F$:
\begin{align}
	\braket{l_j} &= \bra{\psi_\theta}L_j^2\ket{\psi_\theta} = J_{jj} \\
	\Re\braket{l_j}{l_k} &= \Re\bra{\psi_\theta}L_jL_k\ket{\psi_\theta} = J_{jk} \\
	\Im\braket{l_j}{l_k} &= \Im\bra{\psi_\theta}L_jL_k\ket{\psi_\theta} = \tilde{J}_{jk} \\
	\begin{split}
	\braket{o_j}{o_k} &= \bra{\psi}\bra{\xi}\sum_n f_j(n)\ketbra{n} \sum_m f_k(m) \ketbra{m} \ket{\psi}\ket{\xi} = \sum_m f_j(m)f_k(m) \bra{\psi}\bra{\xi}\ketbra{m}\ket{\psi}\ket{\xi} \\
	&= \sum_m p(m) \frac{\partial_jp(m)}{p(m)}\frac{\partial_kp(m)}{p(m)} = F_{jk}
	\end{split} \\
	\begin{split}
	\Re\braket{o_j}{l_k} &= \Re \sum_m \frac{\partial_jp(m)}{p(m)} \bra{\psi}\bra{\xi}\ketbra{m} (L_k\otimes \openone_a)\ket{\psi}\ket{\xi}\\
	 &= \frac{1}{2}\sum_m \frac{\partial_jp(m)}{p(m)}\bra{m}\left\{ (L_k\otimes \openone_a), \ket{\psi}\ketbra{\xi}\bra{\psi} \right\}\ket{m} \\
	 &= \sum_m \frac{\partial_jp(m)}{p(m)} \qtr[\ketbra{m}\partial_k\rho\otimes \ketbra{\xi}]  = \sum_m \frac{\partial_jp(m)\partial_kp(m)}{p(m)} = F_{jk}
	\end{split}
\end{align}
Note that $J$ is diagonal with an appropriate choice of $\ket{l_j}$, and subsequently so is $F$ with an appropriate choice of $\ket{o_j}$. These choices correspond to choosing the canonical parametrisation, as described in the main text. 

The following geometric lemma is key:
\begin{lemma}[Branciard~\cite{branciard_error-tradeoff_2013}]
	Let $\hat{a}$ and $\hat{b}$ be Euclidean unit vectors. For any two orthogonal vectors $\vec{x}$ and $\vec{y}$,
	\begin{equation}
		\|\hat{a}-\vec{x}\|^2 + \|\hat{b}-\vec{y}\|^2 + 2\sqrt{1-\chi^2}\|\hat{a}-\vec{x}\|\|\hat{b}-\vec{y}\|\geq \chi^2,
	\end{equation}
	where $\chi=\hat{a}\cdot\hat{b}$ and $\|\cdot\|$ denotes the norm of a vector. 
\end{lemma}
Let
\begin{equation}
		\hat{a} = \frac{1}{\sqrt{\braket{l_1}}}\begin{pmatrix}
			\Re\ket{l_1} \\
			\Im\ket{l_1}
		\end{pmatrix}, \ \hat{b} = \frac{1}{\sqrt{\braket{l_2}}}\begin{pmatrix}
			\Im\ket{l_2} \\
			-\Re\ket{l_2}
		\end{pmatrix}, \ \vec{x} = \frac{1}{\sqrt{\braket{l_1}}}\begin{pmatrix}
			\Re\ket{o_1} \\
			\Im\ket{o_1}
		\end{pmatrix}, \ \vec{y} = \frac{1}{\sqrt{\braket{l_2}}}\begin{pmatrix}
			\Im\ket{o_2} \\
			-\Re\ket{o_2}
		\end{pmatrix}.
\end{equation}
These vectors satisfy:
\begin{align}
    \|\hat{a}\| &= \hat{a}^\top \hat{a} = \frac{1}{\braket{l_1}}((\Re\ket{l_1})^\top(\Re\ket{l_1}) + (\Im\ket{l_1})^\top(\Im\ket{l_1})) = \frac{1}{\braket{l_1}}\braket{l_1}=1, \\
    \|\hat{b}\| &= 1, \\
    \hat{a}\cdot\hat{b} &= \frac{1}{\sqrt{\braket{l_1}\braket{l_2}}}((\Re\ket{l_1})^\top(\Im\ket{l_2})-(\Im\ket{l_1})^\top(\Re\ket{l_2}) = \frac{\Im\braket{l_1}{l_2}}{\sqrt{\braket{l_1}\braket{l_2}}}, \\
    \vec{x}\cdot\vec{y} &= \frac{\Im\braket{o_1}{o_2}}{\sqrt{\braket{l_1}\braket{l_2}}} = \frac{\Im\bra{\xi}\bra{\psi_\theta}O_1O_2\ket{\psi_\theta}\ket{\xi}}{\sqrt{\braket{l_1}\braket{l_2}}} = \frac{\bra{\xi}\bra{\psi_\theta}[O_1,O_2]\ket{\psi_\theta}\ket{\xi}}{2i\sqrt{\braket{l_1}\braket{l_2}}} = 0,
\end{align}
and thus the conditions in Lemma 1. Furthermore, we have
\begin{equation}
	\|\hat{a}-\vec{x}\|^2 = \frac{1}{\braket{l_1}}(\bra{l_1}-\bra{o_1})(\ket{l_1}-\ket{o_1}), \ \|\hat{b}-\vec{y}\|^2 = \frac{1}{\braket{l_2}}(\bra{l_2}-\bra{o_2})(\ket{l_2}-\ket{o_2}),
\end{equation}
which upon applying Lemma 1 gives
\begin{equation}
	\frac{J_{11}-F_{11}}{J_{11}} + \frac{J_{22}-F_{22}}{J_{22}} + 2\sqrt{1-\chi^2}\sqrt{\frac{J_{11}-F_{11}}{J_{11}}\frac{J_{22}-F_{22}}{J_{22}}} \geq \chi^2 \label{eq:regret}
\end{equation}
with
\begin{equation}
	\chi = \frac{\Im\braket{l_1}{l_2}}{\sqrt{\braket{l_1}\braket{l_2}}} = \frac{\bra{\psi_\theta}[L_1,L_2]\ket{\psi_\theta}}{2i\sqrt{J_{11}J_{22}}}.
\end{equation}

In the canonical parameterisation, where $J=I$ and $F$ is diagonal, Eq.~\eqref{eq:regret} can be rewritten as
\begin{equation}
	2-\tr[F] +2\sqrt{1-\beta^2}\sqrt{1-\tr[F]+\det[F]}\geq \beta^2,
\end{equation}
where $\beta = \bra{\psi_\theta}[L_1,L_2]\ket{\psi_\theta}/(2i)$. 
This inequality can then be rearranged (using $\tr X = 1+\det[X]-\det[I-X]$, valid for $2\times 2$ matrices) to obtain
\begin{equation}
	\sqrt{\det[F]} - \sqrt{\det[I-F]}\leq \sqrt{1-\beta^2},
\end{equation} 
and upon reversing the transformation to the canonical parametrisation, we get Eq.~(6) from the main text (factors of $Q$ drop out due to its orthogonality). Note that in the original parameterisation, $\beta$ is equal to the absolute value of the eigenvalues of $D=J^{-1}\tilde{J}$, which is itself invariant under reparametrisation.

\subsection{\texorpdfstring{$\beta=1$}{b=1} for two-dimensional systems}

We can show that $\beta=1$ for any two-dimensional system. To do this, we work in a basis where the density matrix at the true parameter value is $\rho = \left(\begin{smallmatrix}1 & 0 \\
0 & 0\end{smallmatrix}\right)$. Let the (Hermitian) SLD operators be arbitrary:
\begin{equation}
    L_1 = \begin{pmatrix}
        a_1 & b_1+ic_1 \\
        b_1-ic_1 & d_1
    \end{pmatrix}, \quad L_2 = \begin{pmatrix}
        a_2 & b_2+ic_2 \\
        b_2-ic_2 & d_2
    \end{pmatrix}. 
\end{equation}
The condition $\partial_i(\Tr[\rho])=0$ implies $\Tr[\rho L_i]=0$, which forces $a_i=0$. Then after some straightforward calculation we find:
\begin{equation}
    J = \begin{pmatrix}
        b_1^2+c_1^2 & b_1 b_2 + c_1 c_2 \\
        b_1 b_2 + c_1 c_2 & b_2^2 +c_2^2
    \end{pmatrix}, \quad \tilde{J} =\begin{pmatrix}
        0 & b_2 c_1 - b_1 c_2  \\
        b_1c_2-b_2c_1 & 0 
    \end{pmatrix},
\end{equation}
and the eigenvalues of $J^{-1}\tilde{J}$ are $\pm i$, so $\beta=1$. 

\section{Measurements saturating the Fisher Information inequality}\label{sec:measurements}

Here, we provide constructions of measurements that have the classical Fisher information of any $G$ with equality in Eq.~(6) of the main text. To do this, we work in a parameterisation where $J=I$ (since any problem can be transformed into this parameterisation, as we explicitly show below). We first provide a measurement that gives a diagonal classical Fisher information. Then, we explain how a different classical Fisher information with the same eigenvalues can be obtained by transforming the original measurement. The optimal measurements in the original parameterisation can be obtained by applying the inverse of the  transformation used for reparameterisation to the measurement operators.  

We use different measurements for the two cases $\beta=1$ and $\beta<1$. When $\beta=1$, the system is effectively two-dimensional, and for $\beta<1$ the system is effectively three-dimensional. For reference, we also detail existing measurements that are optimal: for $\beta=1$, the measurement proposed by Wang \textit{et al.}~\cite{wang_tight_2025}, saturates our inequality, as their (different) trade-off relation is equivalent to ours when $\beta=1$. For $\beta<1$, the projective measurement derived by Branciard \cite{branciard_error-tradeoff_2013} is also optimal. 

\subsection{Measurement for \texorpdfstring{$\beta<1$}{b<1}}
For simplicity, we provide an optimal measurement in a basis where the estimation problem (as specified by the probe state and its derivatives) has a simple form. As we outline below, any problem can be recast into this standard form. Subsequently, the optimal measurement can be determined by undoing the effect of the reparameterisation (i.e., applying the inverse of the transformation to the standard form POVM). We begin by demonstrating the transformation to the standard form, where $J=I$.  

Let $\ket{\psi_0}\in \mathcal{H}_d$ be the probe state and $\ket{\psi_1},\ket{\psi_2}\in \mathcal{H}_d$ its derivatives evaluated at the true parameters. We start by representing these in a three-dimensional space (e.g., projecting onto the subspace $\operatorname{Span}\{\ket{\psi_0},\ket{\psi_1},\ket{\psi_2}\}$), choosing a basis such that $\ket{\psi_0} = (1,0,0)^\top$. We also choose the basis such that $\braket{\psi_j}{\psi_0}=0$ for $j=1,2$. Starting from an arbitrary basis, this can be achieved by the transformation $\ket{\psi_j} \rightarrow \ket{\psi_j}-\ket{\psi_0}\braket{\psi_0}{\psi_j}$. Furthermore, we choose the basis such that the real part of $\braket{\psi_1}{\psi_2}$ is zero. This can be achieved by the transformation that diagonalises the real part of the two-by-two matrix with elements $\braket{\psi_j}{\psi_k}$. At this point, we have a basis such that
\begin{equation}
    \braket{\psi_j}{\psi_k} = \begin{pmatrix}
        1 & 0 & 0 \\
        0 & \braket{\psi_1} & ic \\
        0 & -ic & \braket{\psi_2}
    \end{pmatrix}, \quad c = \Im\braket{\psi_1}{\psi_2}.
\end{equation}
Next, we reparameterise the problem such that $\braket{\psi_j}=1/4$ for $j=1,2$. Then, we have a basis such that 
\begin{equation}
    \ket{\psi_0} = \begin{pmatrix}
        1 \\
        0 \\
        0 
    \end{pmatrix}, \quad \ket{\psi_1} = \frac{1}{2}\begin{pmatrix}
        0 \\
        u_1 \\
        u_2
    \end{pmatrix}, \quad \ket{\psi_2} = \frac{1}{2}\begin{pmatrix}
        0 \\
        v_1 \\
        v_2
    \end{pmatrix}, \label{eq:S20}
\end{equation}
with $|u_1|^2+|u_2|^2=|v_1|^2+|v_2|^2=1$. We have that $L_j = 2(\ketbra{\psi_j}{\psi_0}+\ketbra{\psi_0}{\psi_j})$, so
\begin{align}
    i\beta = \bra{\psi_0}L_1L_2\ket{\psi_0} &= 4\bra{\psi_0}\left(\ketbra{\psi_1}{\psi_0}+\ketbra{\psi_0}{\psi_1}\right)\left(\ketbra{\psi_2}{\psi_0}+\ketbra{\psi_0}{\psi_2}\right)\ket{\psi_0} \\
    &= 4\braket{\psi_0}^2\braket{\psi_1}{\psi_2}\\
    &= u_1^*v_1 + u_2^*v_2.
\end{align}
Finally, we perform another basis change with the unitary transformation
\begin{equation}
    U = \begin{pmatrix}
        1 & 0 & 0 \\
        0 & 0 & 0 \\
        0 & 0 & 0
    \end{pmatrix} + 4\left(\ketbra{\varsigma_1}{\psi_1} + \ketbra{\bar{\varsigma}_1}{\bar{\psi}_1}\right),
\end{equation}
where
\begin{align}
    \ket{\varsigma_1} &= \frac{1}{2}\begin{pmatrix}
        0 \\
        i\sin\eta \\
        \cos\eta
    \end{pmatrix} \\
    \ket{\bar{\varsigma}_1} &= \frac{\ket{\varsigma_2}-4\ket{\varsigma_1}\braket{\varsigma_1}{\varsigma_2}}{2\sqrt{1/4-4\left|\braket{\varsigma_1}{\varsigma_2}\right|^2}} \\
    \ket{\bar{\psi}_1} &= \frac{\ket{\psi_2}-4\ket{\psi_1}\braket{\psi_1}{\psi_2}}{2\sqrt{1/4-4\left|\braket{\psi_1}{\psi_2}\right|^2}},
\end{align}
and 
\begin{equation}
    \ket{\varsigma_2} = \frac{1}{2}\begin{pmatrix}
        0 \\
        \cos\eta\\
        -i\sin\eta
    \end{pmatrix}.
\end{equation}
These states are defined such that $\braket{\bar{\varsigma}_1}{\varsigma_1} = \braket{\bar{\psi}_1}{\psi_1}=0$. Applying the unitary transformation to the representation in Eq.~\eqref{eq:S20}, we get
\begin{align}
    \ket{\psi_0} \rightarrow U\ket{\psi_0} &= \ket{\psi_0} \\
    &= \begin{pmatrix}
        1 \\
        0 \\
        0
    \end{pmatrix} \\
    \ket{\psi_1} \rightarrow U\ket{\psi_1} &= 4\braket{\psi_1}\ket{\varsigma_1} \\
    &= \ket{\varsigma_1} \\
    &= \frac{1}{2}\begin{pmatrix}
        0 \\
        i\sin\eta \\
        \cos\eta
    \end{pmatrix} \\
    \ket{\psi_2} \rightarrow U\ket{\psi_2} &= 4\braket{\psi_1}{\psi_2}\ket{\varsigma_1} + 4\braket{\bar{\psi}_1}{\psi_2}\ket{\bar{\varsigma}_1} \\
    &=4\braket{\psi_1}{\psi_2}\ket{\varsigma_1} + 4\frac{\braket{\psi_2}-4\braket{\psi_1}{\psi_2}\braket{\psi_2}{\psi_1}}{2\sqrt{1/4-4\left|\braket{\psi_1}{\psi_2}\right|^2}}\frac{\ket{\varsigma_2}-4\ket{\varsigma_1}\braket{\varsigma_1}{\varsigma_2}}{2\sqrt{1/4-4\left|\braket{\varsigma_1}{\varsigma_2}\right|^2}} \\
    &=4\braket{\psi_1}{\psi_2}\ket{\varsigma_1} + 4\frac{1/4-4\left|\braket{\psi_1}{\psi_2}\right|^2}{2\sqrt{1/4-4\left|\braket{\psi_1}{\psi_2}\right|^2}}\frac{\ket{\varsigma_2}-4\ket{\varsigma_1}\braket{\varsigma_2}{\varsigma_1}}{2\sqrt{1/4-4\left|\braket{\varsigma_1}{\varsigma_2}\right|^2}} \\
    &= i\beta\ket{\varsigma_1} + 4\frac{1/4-\beta^2/4}{2\sqrt{1/4-\beta^2/4}}\frac{\ket{\varsigma_2}-i\beta\ket{\varsigma_1}}{2\sqrt{1/4-\beta^2/4}} \\
    &= i\beta\ket{\varsigma_1} + \ket{\varsigma_2}-i\beta\ket{\varsigma_1} \\
    &= \ket{\varsigma_2} \\
    &= \frac{1}{2}\begin{pmatrix}
        0 \\
        \cos\eta\\
        -i\sin\eta
    \end{pmatrix}
\end{align}
The two-parameter model is now in the standard form described by
\begin{equation}
    \ket{\psi_0} = \begin{pmatrix}
        1 \\
        0 \\
        0 
    \end{pmatrix}, \quad \ket{\psi_1} = \frac{1}{2}\begin{pmatrix}
        0 \\
        i\sin\eta \\
        \cos\eta
    \end{pmatrix}, \quad \ket{\psi_2} = \frac{1}{2}\begin{pmatrix}
        0 \\
        \cos\eta \\
        -i\sin\eta
    \end{pmatrix}. \label{eq:standardform}
\end{equation}
A consequence of this transformation is that any two pure two-parameter models with a particular value of $\beta$ are isomorphic.

\subsubsection{Measurement for diagonal \texorpdfstring{$F$}{F}}
In the standard form, the projective measurement given by the orthonormal vectors
\begin{equation}
    \ket{\pi_1} = \frac{1}{\sqrt{3}}\begin{pmatrix}
        1 \\ e^{i\varphi} \\ e^{i\varphi}
    \end{pmatrix}, \ \ket{\pi_2} = \begin{pmatrix}
        1/\sqrt{3} \\ x_\varphi \\ y_\varphi 
    \end{pmatrix}, \ \ket{\pi_3} = \begin{pmatrix}
        1/\sqrt{3} \\ y_\varphi \\ x_\varphi
    \end{pmatrix}, \label{eq:projectors}
\end{equation}
where $x_\varphi = (3e^{-i\varphi}-\sqrt{3}e^{i\varphi})/6$ and $y_\varphi = -(3e^{-i\varphi}+\sqrt{3}e^{i\varphi})/6$, is optimal. It gives the classical Fisher information
\begin{equation}
    F_{ij}(\varphi) = \sum_k \frac{1}{\Tr[\ketbra{\psi_0}\ketbra{\pi_k}]}\Tr[(\ketbra{\psi_i}{\psi_0}+\ketbra{\psi_0}{\psi_i})\ketbra{\pi_k}]\Tr[(\ketbra{\psi_j}{\psi_0}+\ketbra{\psi_0}{\psi_j})\ketbra{\pi_k}],
\end{equation}
which, after some work, evaluates to
\begin{equation}
    F = \begin{pmatrix}
        \cos^2(\varphi-\eta) & 0 \\
        0 & \cos^2(\varphi+\eta)
    \end{pmatrix}.
\end{equation}
In other words, we have found, for the parametrisation where $J=I$, a POVM which gives rise to a diagonal $F$ which saturates the Fisher Information inequality.
The POVM in the original parametrisation can then be found by applying the inverse of $U$ to the projectors in Eq.~\eqref{eq:projectors}.

\subsubsection{Measurement for arbitrary \texorpdfstring{$F$}{F}}

Working again in the standard form (as in Eq.~\eqref{eq:standardform}), we now show how to modify the measurement to produce a classical Fisher information $QF(\varphi)Q^\top$ for arbitrary orthogonal $Q$. Let us first define the vectors
\begin{equation}
    \begin{pmatrix}
        \ket{\Psi_1} \\
        \ket{\Psi_2}
    \end{pmatrix} = Q\begin{pmatrix}
        \ket{\psi_1} \\
        \ket{\psi_2}
    \end{pmatrix}.
\end{equation}

We next define a unitary transformation
\begin{equation}
    U_Q = \begin{pmatrix}
        1 & 0 & 0 \\
        0 & 0 & 0 \\
        0 & 0 & 0 
    \end{pmatrix} + 4(\ketbra{\Psi_1}{\psi_1} + \ketbra{\zeta}{\bar\zeta}),
\end{equation}
where 
\begin{align}
    \ket{\bar\zeta} &= \frac{\ket{\psi_2}-4\ket{\psi_1}\braket{\psi_1}{\psi_2}}{2\sqrt{1/4-4\left|\braket{\psi_1}{\psi_2}\right|^2}} \\
    \ket{\zeta} &= \frac{\ket{\Psi_2}-4\ket{\Psi_1}\braket{\psi_1}{\psi_2}}{2\sqrt{1/4-4\left|\braket{\psi_1}{\psi_2}\right|^2}}.
\end{align}

We find that $U_Q$ keeps $\ket{\psi_0}$ fixed and maps $\ket{\psi_j}$ to $\ket{\Psi_j}$:
\begin{align}
    U_Q\ket{\psi_1} &= \ket{\Psi_1} + 4\ket{\zeta}\left(\frac{\braket{\psi_2}{\psi_1}-4\braket{\psi_1}\braket{\psi_1}{\psi_2}^*}{2\sqrt{1/4-4\left|\braket{\psi_1}{\psi_2}\right|^2}}\right) = \ket{\Psi_1} \\
    U_Q\ket{\psi_2} &= 4\ket{\Psi_1}\braket{\psi_1}{\psi_2} + 4\frac{\ket{\Psi_2}-4\ket{\Psi_1}\braket{\psi_1}{\psi_2}}{2\sqrt{1/4-4\left|\braket{\psi_1}{\psi_2}\right|^2}}\left(\frac{\braket{\psi_2}{\psi_2}-4\braket{\psi_1}{\psi_2}\braket{\psi_1}{\psi_2}^*}{2\sqrt{1/4-4\left|\braket{\psi_1}{\psi_2}\right|^2}}\right) = \ket{\Psi_2},
\end{align}
using the property that $\braket{\psi_1}{\psi_1}=\braket{\psi_2}{\psi_2}=1/4$.
It can now be shown that the projective measurement given by the orthonormal vectors $U_Q^\dagger \ket{\pi_1}, U_Q^\dagger \ket{\pi_2}, U_Q^\dagger \ket{\pi_3}\}$, has classical Fisher information $\tilde{F}(\varphi) = QF(\varphi)Q^\top$:
\begin{align}
    \tilde{F}(\varphi)_{ij} &= \sum_k \frac{\Tr[(\ketbra{\psi_i}{\psi_0}+\ketbra{\psi_0}{\psi_i})U_Q^\dagger\ketbra{\pi_k}U_Q]\Tr[(\ketbra{\psi_j}{\psi_0}+\ketbra{\psi_0}{\psi_j})U_Q^\dagger\ketbra{\pi_k}U_Q]}{\Tr[\ketbra{\psi_0}U_Q^\dagger\ketbra{\pi_k}U_Q]} \\
    &= \sum_k \frac{\Tr[(U_Q\ketbra{\psi_i}{\psi_0}+\ketbra{\psi_0}{\psi_i}U_Q^\dagger)\ketbra{\pi_k}]\Tr[(U_Q\ketbra{\psi_j}{\psi_0}+\ketbra{\psi_0}{\psi_j}U_Q^\dagger)\ketbra{\pi_k}]}{\Tr[\ketbra{\psi_0}\ketbra{\pi_k}]} \\
    &= \sum_k \frac{1}{\Tr[\ketbra{\psi_0}\ketbra{\pi_k}]}\Tr[(\ketbra{\Psi_i}{\psi_0}+\ketbra{\psi_0}{\Psi_i})\ketbra{\pi_k}]\Tr[(\ketbra{\Psi_j}{\psi_0}+\ketbra{\psi_0}{\Psi_j})\ketbra{\pi_k}] \\
    &= \sum_{l,m}\sum_k Q_{il}Q_{jm}\frac{1}{\Tr[\ketbra{\psi_0}\ketbra{\pi_k}]}\Tr[(\ketbra{\psi_l}{\psi_0}+\ketbra{\psi_0}{\psi_l})\ketbra{\pi_k}]\Tr[(\ketbra{\psi_m}{\psi_0}+\ketbra{\psi_0}{\psi_m})\ketbra{\pi_k}] \\
    &= \sum_{l,m} Q_{il}Q_{jm} F(\varphi)_{lm} \\
    &= [QF(\varphi)Q^\top]_{ij},
\end{align}
where we have used the relationship between $\ket{\Psi_i}$ and $\ket{\psi_j}$:
\begin{equation}
    \ket{\Psi_i} = \sum_l Q_{il}\ket{\psi_l}.
\end{equation}
To summarise, we have now shown that in the parametrisation where $J=I$, we can explicitly construct a measurement for any $F$ saturating the Fisher information inequality. The POVM in the original parametrisation can then be found by applying the inverse of $U_Q^\dagger U$ to the projectors in Eq.~\eqref{eq:projectors}. Therefore, the above constructions allow a measurement to be determined for any $G$ with equality in Eq.~(6) in the main text. 

\subsection{Measurement for \texorpdfstring{$\beta=1$}{b=1}} 

When $\beta=1$, the states $\ket{l_1}$ and $\ket{l_2}$ described by Eq.~(3) are linearly dependent. Therefore, $\operatorname{Span}\{\ket{\psi_0},\ket{\psi_1},\ket{\psi_2}\}$ is two-dimensional. We can thus perform a transformation, similar to above, to put the problem into the standard form
\begin{equation}
    \ket{\psi_0} = \begin{pmatrix}
        1 \\
        0
    \end{pmatrix}, \ \ket{\psi_1} = \frac{1}{2}\begin{pmatrix}
        0 \\
        1
    \end{pmatrix}, \ \ket{\psi_2} = \frac{1}{2}\begin{pmatrix}
        0 \\
        i
    \end{pmatrix},
\end{equation}
where $\braket{\psi_j} = 1/4$ for $j=1,2$ and $\braket{\psi_1}{\psi_2} = i\beta/4 = i/4$. 

In this basis, an optimal POVM is given by the elements
\begin{align}
    \Pi_1 &= \frac{\alpha}{2} \left(\openone + L_1\right) \label{eq:60}\\
    \Pi_2 &= \frac{\alpha}{2} \left(\openone - L_1\right) \\
    \Pi_3 &= \frac{1-\alpha}{2} \left(\openone + L_2\right) \\
    \Pi_4 &= \frac{1-\alpha}{2} \left(\openone - L_2\right) \label{eq:63}
\end{align}
with $L_1 = 2(\ketbra{\psi_1}{\psi_0} + \ketbra{\psi_0}{\psi_1})$ and $L_2= 2(\ketbra{\psi_2}{\psi_0} + \ketbra{\psi_0}{\psi_2})$. This POVM measures the optimal observable for estimating $\theta_1$ with probability $\alpha$, and the optimal observable for estimating $\theta_2$ with probability $1-\alpha$ (i.e., given $N$ probe states, $\approx \alpha N$ are used to estimate $\theta_1$ and $\approx (1-\alpha)N$ are used to estimate $\theta_2$). This gives the classical Fisher information
\begin{equation}
    F = \begin{pmatrix}
        \alpha & 0 \\
        0 & 1-\alpha
    \end{pmatrix},
\end{equation}
which we can put into the desired form by choosing $\alpha = \cos^2(\varphi-\pi/4)$. 

As before, we can transform the POVM to give the Fisher information $QFQ^\top$, using the unitary that implements the transformation $\begin{pmatrix}
    \ket{\psi_1} \\
    \ket{\psi_2}
\end{pmatrix} \rightarrow Q\begin{pmatrix}
    \ket{\psi_1} \\
    \ket{\psi_2}
\end{pmatrix}$. This is equivalent to replacing $L_i$ in Eqs.~\eqref{eq:60}--\eqref{eq:63} with $\tilde{L}_i$ defined by $\begin{pmatrix}
    \tilde L_1 \\
    \tilde L_2
\end{pmatrix} = Q\begin{pmatrix}
    L_1\\
    L_2
\end{pmatrix}$. 

\subsection{Alternative measurement for \texorpdfstring{$\beta<1$}{b<1}}

For $\beta<1$, the measurement presented by Branciard \cite{branciard_error-tradeoff_2013} that saturates Branciard's error trade-off relation, specialised to the case of approximately measuring the two SLD operators, is optimal. For completeness, here we detail the measurement and calculate its Fisher information, of which the off-diagonal term is not relevant to the error trade-off relation, but is important to our work\footnote{We remark that Branciard \cite{branciard_error-tradeoff_2013} also presented a measurement for the case when $\beta=1$ that saturates the error trade-off relation, but it has a singular Fisher information matrix and thus does not saturate our inequality.}.  

Working in the parameterisation where $J=I$, we have 
\begin{align}
	\bra{\psi}L_i\ket{\psi} &= 0, \\
	\bra{\psi}L_i^2\ket{\psi} &= 1, \\
	\bra{\psi}L_1L_2\ket{\psi} &= i\beta, \ \beta\in[0,1).
\end{align}
 
Let $\eta = \arcsin(\beta)/2$, and for an angle $\varphi \in [-|\eta|,|\eta|]$ and $q,r\in\R$, define the complex numbers
\begin{align}
	a &= q\cos(\varphi+\eta)+ir\sin(\varphi+\eta), \\
	b &= r\cos(\varphi-\eta)+iq\sin(\varphi-\eta).
\end{align}
Then, define
\begin{equation}
	\ket{\bar{m}_1} = (\openone + bL_1+aL_2)\ket{\psi}. \label{eq:m1bar}
\end{equation}
Next, define an operator
\begin{equation}
	D = \bra{\bar{m}_1} L_2\ket{\psi}L_1 - \bra{\bar{m}_1}L_1\ket{\psi}L_2, \label{eq:D}
\end{equation}
and set real numbers $s,t$ such that $s\times t=\braket{\bar{m}_1}/(1-\beta^2)$. Then, define 
\begin{align}
	\ket{\bar{m}_2} &= \left(\braket{\bar{m}_1}\openone + sD\right)\ket{\psi} -\ket{\bar{m}_1}, \label{eq:m2bar}\\
	\ket{\bar{m}_3} &= \left(\braket{\bar{m}_1}\openone -tD\right)\ket{\psi} -\ket{\bar{m}_1}. \label{eq:m3bar} 
\end{align}
Branciard showed that $\{\ket{\bar{m}_1},\ket{\bar{m}_2},\ket{\bar{m}_3}\}$ are orthogonal and span $\operatorname{Span}\{\ket{\psi},L_1\ket{\psi},L_2\ket{\psi}\}$. The vectors can thus be normalised to form an orthonormal basis $\{\ket{m_1},\ket{m_2},\ket{m_3}\}$ of $\operatorname{Span}\{\ket{\psi},L_1\ket{\psi},L_2\ket{\psi}\}$. Where necessary, the basis can be extended to an orthonormal basis $\{\ket{m_1},\ket{m_2},\ket{m_3},\dots,\ket{m_d}\}$ of $\mathcal{H}_d$. The projective measurement $\{\ketbra{m_i}\}$ is then optimal, in the sense of saturating the inequality, as we shall now show. 

The outcome probabilities are
\begin{align}
	p(1) &= |\braket{\psi}{m_1}|^2 = \frac{1}{\braket{\bar{m}_1}} \\
	p(2) &= |\braket{\psi}{m_2}|^2 = \frac{|\braket{\bar{m}_1}-1|^2}{\braket{\bar{m}_2}} \\
	p(3) &= |\braket{\psi}{m_3}|^2 = \frac{|\braket{\bar{m}_1}-1|^2}{\braket{\bar{m}_3}},
\end{align}
and the classical Fisher information is
\begin{align}
	F_{jk} &= \sum_{i=1}^3 \frac{1}{p(i)}\frac{\partial p(i)}{\partial\theta_j} \frac{\partial p(i)}{\partial\theta_k},\\
	&= \sum_{i=1}^3 \frac{1}{p(i)}\left(\bra{m_i}\left[\ketbra{\partial_j\psi}{\psi}+\ketbra{\psi}{\partial_j\psi}\right]\ket{m_i}\right) \left(\bra{m_i}\left[\ketbra{\partial_k\psi}{\psi}+\ketbra{\psi}{\partial_k\psi}\right]\ket{m_i}\right),\\
	&= \sum_{i=1}^3 \frac{1}{p(i)} \left(\frac{1}{2}\bra{m_i}L_j\ket{m_i}\right)\left(\frac{1}{2}\bra{m_i}L_k\ket{m_i}\right).
\end{align}
The remaining terms are
\begin{align}
	\frac{1}{2}\bra{m_1}L_1\ket{m_1} &= \frac{1}{2\braket{\bar{m}_1}}\bra{\bar{m}_1}L_1\ket{\bar{m}_1} \\
	&= \frac{1}{\braket{\bar{m}_1}}(\Re(b)-\beta \Im(a)) = \frac{r}{\braket{\bar{m}_1}}(\cos(\varphi-\eta)-\beta \sin(\varphi+\eta)), \\
	\frac{1}{2}\bra{m_1}L_2\ket{m_1} &= \frac{1}{2\braket{\bar{m}_1}}\bra{\bar{m}_1}L_2\ket{\bar{m}_1} \\
	&= \frac{1}{\braket{\bar{m}_1}}(\Re(a)+\beta \Im(b)) =\frac{q}{\braket{\bar{m}_1}}(\cos(\varphi+\eta) + \beta \sin(\varphi-\eta) \\
	\frac{1}{2}\bra{m_2}L_1\ket{m_2} &= \frac{1}{2\braket{\bar{m}_2}}\bra{\bar{m}_2}L_1\ket{\bar{m}_2} = \frac{1}{\braket{\bar{m}_2}}(\braket{\bar{m}_1}-1)(s(1-\beta^2)\Re(a) + \beta \Im(a)-\Re(b)) \\
	\frac{1}{2}\bra{m_2}L_2\ket{m_2} &= \frac{1}{2\braket{\bar{m}_2}}\bra{\bar{m}_2}L_2\ket{\bar{m}_2}  = \frac{1}{\braket{\bar{m}_2}}(1-\braket{\bar{m}_1})(s(1-\beta^2)\Re(b) + \beta \Im(b)+\Re(a)),
\end{align}
and the inner products for $\ket{m_3}$ are the same as for $\ket{m_2}$ except with $2\leftrightarrow 3$ and $s\leftrightarrow -t$. The above results make use of the following inner products
\begin{align}
	\braket{\bar{m}_1} &= 1+|a|^2+|b|^2-2\beta \Im(ab^*) = 1+(1-\beta^2)(q^2+r^2)\\
	\braket{\bar{m}_2} &= (\braket{\bar{m}_1}-1)(\braket{\bar{m}_1}+s^2(1-\beta^2)) \\
	\braket{\bar{m}_3} &= (\braket{\bar{m}_1}-1)(\braket{\bar{m}_1}+t^2(1-\beta^2))\\
	\braket{\bar{m}_1}{\psi} &= 1 \\
	\bra{\psi}D\ket{\psi} &= \bra{\bar{m}_1}D\ket{\psi}=0 \\
	\bra{\bar{m}_1}L_1\ket{\psi} &= b^*-i\beta a^* \\
	\bra{\bar{m}_1}L_2\ket{\psi} &= i\beta b^*+a^* \\
	\bra{\psi}L_1D\ket{\psi} &= \bra{\bar{m}_1}L_1D\ket{\psi}= (1-\beta^2)a^* \\ 
	\bra{\psi}L_2D\ket{\psi} &= \bra{\bar{m}_1}L_2D\ket{\psi}=-(1-\beta^2)b^* \\
	\bra{\psi}D^\dagger D\ket{\psi} &= (1-\beta^2)(\braket{\bar{m}_1}-1)
\end{align}

Finally, the classical Fisher information matrix can be calculated and is independent of $q,r,s,t$:
\begin{equation}
	F = \begin{pmatrix}
		\cos^2(\varphi+\eta) & 0 \\
		0 & \cos^2(\varphi-\eta) 
	\end{pmatrix}.
\end{equation}

\subsection{Alternative measurement for \texorpdfstring{$\beta=1$}{b=1}}
When $\beta=1$, the vectors $\ket{l_1}$ and $\ket{l_2}$ are linearly dependent. Wang \textit{et al}.~constructed an optimal measurement~\cite{wang_tight_2025}. By choosing another vector $\ket{l_\perp}$ such that $\braket{l_\perp}{l_1}=0$, $\braket{l_\perp}{\psi_\theta}\ket{\xi}=0$, $\braket{l_\perp}=1$, the following measurement vectors are optimal \cite{wang_tight_2025}: 
\begin{align}
    \ket{o_1} &= \frac{1}{2}(1-\sin(2\varphi)\ket{l_1} + \frac{i}{2}\cos(2\varphi)\ket{l_\perp} \\
    \ket{o_2} &= \frac{i}{2}(1+\sin(2\varphi)\ket{l_1} + \frac{1}{2}\cos(2\varphi)\ket{l_\perp}
\end{align}
where $\varphi \in[-\pi/4,\pi/4]$. This measurement is optimal because:
\begin{equation}
    F = \begin{pmatrix}
        \braket{o_1}{o_1} & \braket{o_1}{o_2} \\
        \braket{o_2}{o_1} & \braket{o_2}{o_2}
    \end{pmatrix} = \begin{pmatrix}
        \frac{1}{2}(1-\sin(2\varphi)) & 0 \\
        0 & \frac{1}{2}(1+\sin(2\varphi))
    \end{pmatrix} = \begin{pmatrix}
        \cos^2(\varphi  + \pi/4) & 0 \\
        0 & \cos^2(\varphi-\pi/4)
    \end{pmatrix}.
\end{equation}
It then remains to determine the measurement basis $\{\ket{m}\}$, which Wang \textit{et al}.~describe \cite{wang_tight_2025}.  

\section{The Most Informative Cram\'er--Rao bound}

In the main text, we show that the most informative Cram\'er--Rao bound for two-parameter estimation with pure state probes is given by
\begin{equation}
	\mathcal{C}^\text{MI}(W) = \min_{\varphi\in[-\eta,\eta]}\left\{\frac{s_1}{\cos^2(\varphi-\eta)} + \frac{s_2}{\cos^2(\varphi+\eta)} \right\}, \label{eq:CMI}
\end{equation}
where $s_j$ are the eigenvalues of $S=J^{-1/2}WJ^{-1/2}$ and $\eta = \arcsin(\beta)/2$. 

\subsection{Reduction to quartic equation}
Here, we demonstrate that the solution to the minimisation problem over $\varphi$ can be reduced to the solution of a quartic equation in $x=\tan\varphi$, $|x|\leq \tan\eta$. In terms of $x$, the objective is
\begin{equation}
	\frac{s_1}{\cos^2(\varphi-\eta)}+\frac{s_2}{\cos^2(\varphi+\eta)} = (1+x^2)\left(\frac{s_1}{(\cos\eta +x\sin\eta)^2}+\frac{s_2}{(\cos\eta -x\sin\eta)^2}\right)
\end{equation}
Its minimum occurs at a stationary point (as we know that it does not occur at the boundary):
\begin{align}
	0&=\frac{\partial}{\partial x}\left((1+x^2)\left(\frac{s_1}{(\cos\eta +x\sin\eta)^2}+\frac{s_2}{(\cos\eta -x\sin\eta)^2}\right)\right) \\
	&= -(1+x^2)\left(\frac{2s_1\sin\eta}{(\cos\eta+x\sin\eta)^3}-\frac{2s_2\sin\eta}{(\cos\eta-x\sin\eta)^3}\right)+2x\left(\frac{s_1}{(\cos\eta+x\sin\eta)^2}+\frac{s_2}{(\cos\eta-x\sin\eta)^2}\right)
\end{align}
Multiplying both sides by $(\cos\eta+x\sin\eta)^3(\cos\eta-x\sin\eta)^3$ and rearranging gives the fourth-order polynomial
\begin{multline}
	0=2(s_2-s_1)\cos\eta\sin^3\eta \ x^4 + 2(s_1+s_2)(3\cos^2\eta\sin^2\eta+\sin^4\eta)x^3 \\ -6(s_1-s_2)(\cos^3\eta\sin\eta + \cos\eta\sin^3\eta)x^2
	+2(s_1+s_2)(\cos^4\eta + 3\cos^2\eta\sin^2\eta)x - 2(s_1-s_2)\cos^3\eta\sin\eta.
\end{multline}
That is, the solution to the minimisation problem is one of the roots of the above polynomial. Note that $(\cos\eta+x\sin\eta)^3(\cos\eta-x\sin\eta)^3$ is non-zero when $\beta<1$, and when $\beta=1$ it is only zero when $x=\pm \tan\eta=\pm 1$. It is evident by inspecting the objective that the minimum does not occur at these values, unless $s_1=0$ or $s_2=0$ (and in those cases the minimisation is trivial). 

\subsection{Special cases of \texorpdfstring{$\mathcal{C}^\text{MI}$}{CMI}}
\subsubsection{\texorpdfstring{$\beta=0$}{b=0}}
For $\beta=0$, we have $\eta=0$ so the minimum occurs at $\varphi=0$. In this case, $\mathcal{C}^\text{MI}$ coincides with the SLD Cram\'er--Rao bound. 
\begin{equation}
	\mathcal{C}^\text{MI}(W;\beta=0)=s_1+s_2= \tr[S] = \tr[J^{-1/2}WJ^{-1/2}] = \tr[WJ^{-1}] = \mathcal{C}^\text{SLD}(W). 
\end{equation} 
Note that this is the expected result because $\beta=0$ means that the SLD operators are weakly commuting ($\Tr[\rho[L_1,L_2]]=0$) and the SLD bound is attainable for pure states. 

\subsubsection{\texorpdfstring{$\beta=1$}{b=1}}
For $\beta=1$, we have $\eta = \pi/4$. Direct minimisation by setting the derivative 
\begin{equation}
	\frac{\mathrm{d}}{\mathrm{d}\varphi} \left(\frac{s_1}{\cos^2(\varphi-\pi/4)}+\frac{s_2}{\cos^2(\varphi+\pi/4)}\right)=0,
\end{equation}
results in
\begin{equation}
	\mathcal{C}^\text{MI}(W;\beta=1) = (\sqrt{s_1}+\sqrt{s_2})^2 = \left(\tr[\sqrt{J^{-1/2}WJ^{-1/2}}]\right)^2.
\end{equation}
Furthermore, for $\dim\mathcal{H}=2$, we always have $\beta=1$, and we recover Gill and Massar's result, which is also equivalent to Nagaoka's. 

\subsubsection{\texorpdfstring{$s_1=s_2$}{s1=s2}}
The case $s_1=s_2=s$ occurs when the weight is proportional to $J$. In this case we have
\begin{equation}
	\mathcal{C}^\text{MI}(W;W\propto J) = s\min_\varphi \left\{ \frac{1}{\cos^2(\varphi-\eta)}+\frac{1}{\cos^2(\varphi+\eta)} \ \Big\vert \ |\varphi| \leq \eta =\frac{\arcsin{(\beta)}}{2}  \right\}.
\end{equation}
Here, it is clear from Fig.~1 in the main text that the minimum occurs when $\varphi=0$, where the eigenvalues of $G$ are equal. Thus we have 
\begin{equation}
	\mathcal{C}^\text{MI}(W;W\propto J) = \frac{4s}{1+\sqrt{1-\beta^2}},
\end{equation}  
which recovers Matsumoto's result. In particular, with $W=J$, $s=1$. 

\subsubsection{Singular \texorpdfstring{$W$}{W}}

A singular weight $W$ corresponds to estimating a single function of the parameters. In this case, $G=J^{-1/2}WJ^{-1/2}$ is also singular, so $s_2=0$. We then have 
\begin{equation}
	\mathcal{C}^\text{MI}(W;s_2=0) = \min_\varphi \left\{\frac{s_1}{\cos^2(\varphi-\eta)} \ \Big\vert \ |\varphi|\leq \eta = \frac{\arcsin{(\beta)}}{2} \right\} = s_1,
\end{equation}
since the minimum occurs when $\cos^2(\varphi-\eta)$ is maximised at $\varphi=\eta$. A singular $W$ has the form $W=u^\top u$ with $u=(a,b)^\top$, in which case $s_1=u^\top J^{-1}u$. 

\section{Lower bound for mixed states}

Here, we detail the lower bound for mixed states. The derivation is based on a purification that does not increase the quantum Fisher information. Let $\mathcal{C}^\ast(\rho_\theta)$ denote Eq.~\eqref{eq:CMI} with $J$ and $\beta$ calculated from the matrix elements $\qtr[\rho_\theta L_iL_j]$.

Given a mixed state $\rho_\theta\in \mathcal{H}$ that depends on two parameters, let its spectral decomposition be $\rho_\theta = \sum_i p_{i,\theta}\ketbra{\psi_{i,\theta}}$, where $p_i$ and $\ket{\psi}$ may depend on the parameters. We then consider the purification at the local parameter value $\ket{\Psi_{\theta_0}} = \sum_i \sqrt{p_{i,\theta_0}}\ket{\psi_{i,\theta_0}}\otimes \ket{\eta_i}$, where $\{\ket{\eta_i}\}$ is an orthonormal basis for an ancillary (environment) Hilbert space $\mathcal{H}^\text{e}$. This has the property that $\qtr_\text{e}[\ketbra{\Psi_{\theta_0}}]=\rho_{\theta_0}$, where $\qtr_\text{e}[\cdot]$ denotes the partial trace over the environement.

Next, we define the derivatives of the purification by $\ket{\partial_j\Psi}|_{\theta=\theta_0} = \frac{1}{2}(L_j\otimes \openone_e)\ket{\Psi_{\theta_0}}$, where $L_j$ is the SLD operator for $\rho_\theta$ with respect to $\theta_j$~\cite{fujiwara_fibre_2008,kolodynski_efficient_2013,kolodynski_precision_2014}. Hereafter we drop the explicit evaluation at $\theta=\theta_0$ and simply write $\ket{\psi_\theta}$, $\ket{\Psi}$, $\ket{\partial_i \Psi}$, when it is unambiguous. The $\theta$-dependent pure state $\ket{\Psi_\theta}$ that is locally defined by these derivatives has the property that $\qtr_\text{e}[\partial_j(\ketbra{\Psi_\theta})] = \partial_j \rho_\theta$, and the purification therefore faithfully represents the parameterised mixed state. This is shown by:
\begin{align}
    \qtr_\text{e}[\partial_i(\ketbra{\Psi_\theta})] &= \qtr_\text{e}[\ketbra{\partial_i\Psi_\theta}{\Psi_\theta}+ \ketbra{\Psi_\theta}{\partial_i\Psi_\theta}] \\
    &= \frac{1}{2}\qtr_\text{e}[(L_i\otimes \openone_e)\ketbra{\Psi}+ \ketbra{\Psi}(L_i\otimes \openone_e)] \\
    &= \frac{1}{2}\qtr_\text{e}\left[\sum_{jk}\sqrt{p_j}\sqrt{p_k}\left(L_i\ketbra{\psi_j}{\psi_k}\otimes \ketbra{\eta_j}{\eta_k} + \ketbra{\psi_j}{\psi_k}L_i\otimes \ketbra{\eta_j}{\eta_k}\right)\right] \\
    &= \frac{1}{2}\sum_{jk}\sqrt{p_j}\sqrt{p_k}\left(L_i\ketbra{\psi_j}{\psi_k} + \ketbra{\psi_j}{\psi_k}L_i\right)\delta_{jk} \\
    &= \frac{1}{2}(L_i\rho + \rho L_i) \\
    &= \partial_i \rho.
\end{align}

Now, the Cram\'er--Rao bound can be calculated for the purified state, $\mathcal{C}^\text{MI}(\ket{\Psi_\theta})$, which depends on $J(\ket{\Psi_\theta})$ and $\tilde{J}(\ket{\Psi_\theta})$. For the purified state we have the SLD operators
\begin{equation}
    \mathcal{L}_i = 2\left(\ketbra{\Psi}{\partial_i\Psi} + \ketbra{\partial_i\Psi}{\Psi}\right) = \ketbra{\Psi}(L_i\otimes \openone_e) + (L_i\otimes \openone_e)\ketbra{\Psi},
\end{equation}
so the constituents of $J(\ket{\Psi_\theta})$ and $\tilde{J}(\ket{\Psi_\theta})$ are
\begin{align}
    \bra{\Psi}\mathcal{L}_i\mathcal{L}_j\ket{\Psi} &= \bra{\Psi}\left(\ketbra{\Psi}(L_i\otimes \openone_e) + (L_i\otimes \openone_e)\ketbra{\Psi}\right)\left(\ketbra{\Psi}(L_j\otimes \openone_e) + (L_j\otimes \openone_e)\ketbra{\Psi}\right)\ket{\Psi} \\
    &= \bra{\Psi}\Big(\ketbra{\Psi}(L_i\otimes \openone_e)\ketbra{\Psi}(L_j\otimes \openone_e) +\ketbra{\Psi}(L_i\otimes \openone_e)(L_j\otimes \openone_e)\ketbra{\Psi} \\
    &\quad +(L_i\otimes \openone_e)\ketbra{\Psi}    \ketbra{\Psi}(L_j\otimes \openone_e) + (L_i\otimes \openone_e)\ketbra{\Psi}(L_j\otimes \openone_e)\ketbra{\Psi}\Big)\ket{\Psi} \\
    &= 3\underbrace{\bra{\Psi}L_i\otimes \openone_e\ket{\Psi}\bra{\Psi}L_j\otimes \openone_e\ket{\Psi}}_{=0} + \bra{\Psi}L_iL_j\otimes \openone_e\ket{\Psi} \label{eq:zeroterm} \\
    &= \sum_{kl}\sqrt{p_k}\sqrt{p_l}\bra{\psi_k}\otimes\bra{\eta_k}(L_iL_j\otimes \openone_e)\ket{\psi_l}\otimes\ket{\eta_l} \\
    &= \sum_{kl}\sqrt{p_k}\sqrt{p_l}\bra{\psi_k}L_iL_j\ket{\psi_l}\delta_{kl} \\
    &= \sum_{k}p_k \bra{\psi_k}L_iL_j\ket{\psi_k} \\
    &= \qtr[\rho L_iL_j]
\end{align}
where the term in Eq.~\eqref{eq:zeroterm} is zero because $\bra{\Psi}L_i\otimes \openone_e\ket{\Psi}=\qtr[\rho L_i] = \frac{1}{2}\qtr[\rho L_i + L_i\rho]=\qtr[\partial_i\rho] = \partial_i\qtr[\rho]=0$. This means that $J(\ket{\Psi_\theta})=J(\rho_\theta)$ and $\tilde{J}(\ket{\Psi_\theta})=\tilde{J}(\rho_\theta)$, and subsequently $\beta(\ket{\Psi_\theta}) = \beta(\rho_\theta)$. Therefore, $\mathcal{C}^\text{MI}(\ket{\Psi_\theta}) = \mathcal{C}^\ast(\rho_\theta)$ is a lower bound for the mean squared error of $\rho_\theta$ (for the same reason that $\mathcal{C}^\text{MI}$ is a lower bound for the mean squared error of $\ket{\Psi_\theta}$). Unlike in the pure state case, the lower bound is not in general attainable because the optimal measurement may act on the environment space, which an experimenter does not have access to. 

\section{Details on displacement sensing with grid states}

As discussed in Ref.~\cite{duivenvoorden_single-mode_2017}, grid states are well-suited to simultaneously estimating displacements in phase space. The idea is that a grid state is an approximate simultaneous eigenstate of the modular position and momentum ($\hat{q}$, $\hat{p}$, up to $\sqrt{2\pi}$). Ref.~\cite{duivenvoorden_single-mode_2017} derived the lower bound for the sum of the mean squared errors of displacement estimates: $V(u)+V(v) \geq 1/(2\langle\hat{n}\rangle+1)$, where $\langle\hat{n}\rangle$ is the mean photon number (see below for the definition of the displacements). Here, we derive the QFI and $\mathcal{C}^\text{MI}$ for this problem. 

We consider a unitary displacement operation $D(u,v) = e^{-iu\hat{p}+iv\hat{q}}$ (with true values $u=v=0$), applied to a probe grid state defined by
\begin{equation}
	\ket{\psi^\text{grid}}=\mathcal{N}_\Delta\left( \frac{2}{\pi} \right)^{1/4}\sum_{t=-\infty}^\infty e^{-\pi\Delta^2t^2}\int \mathrm{d}q~e^{-(q-\sqrt{2\pi}t)^2/(2\Delta)^2}\ket{q}, 
\end{equation}
where $\Delta$ parameterises the squeezing and $\mathcal{N}_\Delta$ is a normalisation factor (which approaches 1 as $\Delta \rightarrow 0$). Equivalently, the state in the $p$-quadrature basis is
\begin{equation}
    \ket{\psi^\text{grid}} = \mathcal{N}_\Delta\left(\frac{2}{\pi}\right)^{1/4} \sum_{t=-\infty}^\infty \int\mathrm{d}p~e^{-\Delta^2p^2/2}e^{-(p-\sqrt{2\pi}t)^2/(2\Delta^2)}\ket{p}.
\end{equation}

We wish to calculate the matrix elements
\begin{align}
	J_{ij} &= 4\re[\braket{\partial_i\psi}{\partial_j\psi}-\braket{\partial_i\psi}{\psi}\braket{\psi}{\partial_j\psi}], \\
	\tilde{J}_{ij} &= 4\Im[\braket{\partial_i\psi}{\partial_j\psi}-\braket{\partial_i\psi}{\psi}\braket{\psi}{\partial_j\psi}],
\end{align}
where the derivatives of the probe state at $u=v=0$ are
\begin{equation}
    \ket{\partial_u\psi} = -i\hat{p}\ket{\psi}, \quad \ket{\partial_v\psi} = i\hat{q}\ket{\psi}. 
\end{equation}
From the symmetry of the probe state in the $p$-quadrature basis, $\braket{\partial_u\psi}{\psi} = i\bra{\psi}\hat{p}\ket{\psi} = 0$. Similarly, $\braket{\partial_v\psi}{\psi}=0$. For the off-diagonal elements, we have
\begin{align}
    \braket{\partial_u\psi}{\partial_v\psi} &= -\bra{\psi}\hat{p}\hat{q}\ket{\psi}.
\end{align}
The invariance of the grid state under the transformation $\hat{p}\rightarrow \hat{q}$ and $\hat{q}\rightarrow -\hat{p}$ ~\cite{duivenvoorden_single-mode_2017} means that $\bra{\psi}\hat{p}\hat{q}\ket{\psi}= -\bra{\psi}\hat{q}\hat{p}\ket{\psi}$. Furthermore, 
\begin{align}
    \bra{\psi}\hat{p}\hat{q}\ket{\psi}&= \bra{\psi}(\hat{p}\hat{q}-\hat{q}\hat{p}+\hat{q}\hat{p})\ket{\psi} \\
    &= \bra{\psi}([\hat{p},\hat{q}]+\hat{q}\hat{p})\ket{\psi}\\
    &=-i\braket{\psi} +\bra{\psi}\hat{q}\hat{p}\ket{\psi} \\
    &= -i +\bra{\psi}\hat{p}\hat{q}\ket{\psi}^\ast.
\end{align}
Together, these mean that $\bra{\psi}\hat{p}\hat{q}\ket{\psi} = -i/2$. Therefore $J$ is diagonal,
\begin{equation}
    J = \begin{pmatrix}
        4\bra{\psi}\hat{p}^2\ket{\psi} & 0 \\
        0 & 4\bra{\psi}\hat{q}^2\ket{\psi} 
    \end{pmatrix}
\end{equation}
and 
\begin{equation}
    \tilde{J}  =\begin{pmatrix}
        0 & 2 \\
        -2 & 0 
    \end{pmatrix}.
\end{equation}

For the diagonal elements of $J$, Ref.~\cite{duivenvoorden_single-mode_2017} recognised the property of the trace:
\begin{align}
    \tr[J] &= 4\left(\bra{\psi}\hat{p}^2\ket{\psi} +\bra{\psi}\hat{q}^2\ket{\psi} \right) \\
    &= 4\bra{\psi}(\hat{p}^2+\hat{q}^2)\ket{\psi} \\
    &= 4\bra{\psi}(2\hat{n}+1)\ket{\psi} \\
    &= 4(2\langle\hat{n}\rangle +1).
\end{align}
However, this does not tell us about the individual diagonal elements. Instead, we work in the $q$-quadrature basis.

The action of the displacement on a $q$-quadrature wavefunction $\ket{\psi} = \int \mathrm{d}q~\psi(q)\ket{q}$ is
\begin{equation}
	D(u,v)\ket{\psi} = e^{-iu\hat{p}+iv\hat{q}}\int \mathrm{d}q~\psi(q)\ket{q} = e^{-iuv/2}\int\mathrm{d}q~e^{ivq}\psi(q-u)\ket{q}.
\end{equation}
Therefore, the displaced grid state is
\begin{equation}
	\ket{\psi^\text{grid}_{u,v}} = \mathcal{N}_\Delta e^{-iuv/2}\left( \frac{2}{\pi} \right)^{1/4}\sum_{t=-\infty}^\infty e^{-\pi\Delta^2t^2}\int \mathrm{d}q~e^{ivq}e^{-(q-u-\sqrt{2\pi}t)^2/(2\Delta)^2}\ket{q}.
\end{equation}
Its derivatives (at $u=v=0$) are
\begin{align}
	\ket{\partial_u\psi} &= \mathcal{N}_\Delta\left(\frac{2}{\pi} \right)^{1/4} \frac{1}{\Delta^2}\sum_{t=-\infty}^\infty e^{-\pi\Delta^2t^2}\int \mathrm{d}q~(q-\sqrt{2\pi}t)e^{-(q-\sqrt{2\pi}t)^2/(2\Delta)^2}\ket{q} \\
	\ket{\partial_v\psi} &= i\mathcal{N}_\Delta\left(\frac{2}{\pi} \right)^{1/4} \sum_{t=-\infty}^\infty e^{-\pi\Delta^2t^2}\int \mathrm{d}q~qe^{-(q-\sqrt{2\pi}t)^2/(2\Delta)^2}\ket{q}
\end{align}

It is useful to express the state and its derivatives as
\begin{equation}
	\ket{\psi} = \mathcal{N}_\Delta\sum_{t=-\infty}^\infty \ket{\psi_t},\quad \ket{\partial_i\psi} = \mathcal{N}_\Delta\sum_{t=-\infty}^\infty \ket{\partial_i\psi_t},
\end{equation}
with 
\begin{align}
	\ket{\psi_t} &= \left( \frac{2}{\pi} \right)^{1/4}e^{-\pi\Delta^2t^2}\int \mathrm{d}q~e^{-(q-\sqrt{2\pi}t)^2/(2\Delta)^2}\ket{q} \\
	\ket{\partial_u\psi_t} &= \left(\frac{2}{\pi} \right)^{1/4} \frac{1}{\Delta^2} e^{-\pi\Delta^2t^2}\int \mathrm{d}q~(q-\sqrt{2\pi}t)e^{-(q-\sqrt{2\pi}t)^2/(2\Delta)^2}\ket{q}\\
	\ket{\partial_v\psi_t} &= i\left(\frac{2}{\pi} \right)^{1/4} e^{-\pi\Delta^2t^2}\int \mathrm{d}q~qe^{-(q-\sqrt{2\pi}t)^2/(2\Delta)^2}\ket{q}.
\end{align}

Then we can simplify inner products, e.g.,
\begin{align}
	\braket{\psi} &= \mathcal{N}_\Delta^2\left(\sum_{t_1=-\infty}^\infty \bra{\psi_{t_1}}\right)\left(\sum_{t_2=-\infty}^\infty \ket{\psi_{t_2}}\right) \\
	&= \mathcal{N}_\Delta^2\sum_{t_1=-\infty}^\infty \sum_{t_2=-\infty}^\infty \braket{\psi_{t_1}}{\psi_{t_2}},
\end{align}
which defines $\mathcal{N}_\Delta$.

We have
\begin{align}
	\braket{\partial_u\psi_{t_1}}{\partial_u\psi_{t_2}} &= \left(\frac{2}{\pi}\right)^{1/2}\frac{1}{\Delta^4}e^{-\pi\Delta^2(t_1^2+t_2^2)}\\
	&\int \mathrm{d}q~(q-\sqrt{2\pi}t_1)(q-\sqrt{2\pi}t_2)e^{-(q-\sqrt{2\pi}t_1)^2/(2\Delta^2)}e^{-(q-\sqrt{2\pi}t_2)^2/(2\Delta^2)}dq \\
	&= \frac{\Delta^2-\pi(t_1-t_2)^2}{\sqrt{2}\Delta^3}e^{-\frac{\pi}{2\Delta^2}\left(2\Delta^4(t_1^2+t_2^2)+(t_1-t_2)^2\right)}
\end{align}
which can be summed over $t_1$ and $t_2$ numerically. Similarly, 
\begin{align}
	\braket{\partial_v\psi_{t_1}}{\partial_v\psi_{t_2}} &= \left(\frac{2}{\pi}\right)^{1/2}e^{-\pi\Delta^2(t_1^2+t_2^2)}\int \mathrm{d}q~q^2e^{-(q-\sqrt{2\pi}t_1)^2/(2\Delta^2)}e^{-(q-\sqrt{2\pi}t_2)^2/(2\Delta^2)}dq \\
	&= \frac{\Delta}{\sqrt{2}}\left(\Delta^2+\pi(t_1+t_2)^2\right)e^{-\frac{\pi}{2\Delta^2}\left(2\Delta^4(t_1^2+t_2^2)+(t_1-t_2)^2\right)}.
\end{align}
Numerically, we find that the diagonal elements of $J$ are almost equal. To produce Fig.~2 in the main text, we calculate $J$ and $\langle\hat{n}\rangle$ numerically for a range of values for $\Delta$. For each $\Delta$, we calculate $\beta$, and perform the minimisation to calculate $\mathcal{C}^\text{MI}$.

\bibliography{bib.bib}